# Thickness dependence of superconductivity and superconductor-insulator transition in ultrathin FeSe films on SrTiO$_3$(001) substrate

**Qingyan Wang[1,2], Wenhao Zhang[3], Zuocheng Zhang[3], Yi Sun[1,2], Ying Xing[1,2], Yayu Wang[2,3], Lili Wang[2,3], Xucun Ma[2,3], Qi-Kun Xue[2,3] and Jian Wang[1,2]**

[1]International Center for Quantum Materials, School of Physics, Peking University, Beijing 100871, China

[2]Collaborative Innovation Center of Quantum Matter, Beijing 100871, China

[3]State Key Laboratory of Low-Dimensional Quantum Physics, Department of Physics, Tsinghua University, Beijing 100084, China

E-mail: liliwang@mail.tsinghua.edu.cn; jianwangphysics@pku.edu.cn

**Abstract**

Interface-enhanced high-temperature superconductivity in one unit-cell (UC) FeSe film on SrTiO$_3$(001) (STO) substrate has recently attracted much attention in condensed matter physics and material science. Here, by *ex situ* transport measurements, we report on the superconductivity in FeSe ultra-thin films with different thickness on STO substrate. We find that the onset superconducting transition temperature ($T_c$) decreases with increasing film thickness of FeSe, which is opposite to the behavior usually observed in traditional superconductor films. By systematic post-annealing of 5 UC FeSe films, we observe an insulator to superconductor transition, which is accompanied with a sign change of the dominated charge carriers from holes to electrons at low temperatures according to the corresponding Hall measurement.

**Introduction**

Interface-induced high-temperature superconductivity has recently been observed [1, 2] in the single unit-cell (UC) FeSe film grown on SrTiO$_3$(001) (STO) substrate by molecular beam epitaxy (MBE) technique. This discovery has generated an impressive amount of experimental and theoretical studies [1-20]. It represents the thinnest system of high-temperature superconductivity [21] and the superconductivity is extremely enhanced compared with bulk FeSe [22, 23]. According to *in situ* scanning tunneling microscopy/spectroscopy (STM/STS) measurements, no signature of superconductivity down to 0.4 K is detected on the surface of 2 UC and thicker FeSe films on STO substrate [1, 24]. However, *ex situ* transport measurement shows superconductivity for 1 to 5 UC FeSe films [1, 2, 25-27]. This opposite observation by STM/STS and transport measurments may suggest that the enhanced superconductivity only occurs at the first UC of FeSe on STO since the first UC of FeSe can contribute the transport result but STM/STS detects top surface property of the film. To further understand the origin of interface-enhanced superconductivity, a study of thickness-dependent superconductivity in ultrathin FeSe films on STO substrate is necessary.

We perform *ex situ* transport measurements on the FeSe films with different thickness up to 50 UC grown on STO substrate. Interestingly, the superconductivity is suppressed with increasing thickness, which is opposite to the expectation in superconducting films. Furthermore, by post-annealing of the 5 UC FeSe films, an insulator to superconductor transition is identified by transport measurments.

**Results and discussion**

The FeSe films were grown on insulating STO substrates by MBE. After sufficient annealing in ultra-high vacuum (UHV) system, a series of superconducting FeSe films with different thickness were prepared, and then covered by amorphous Si layers for *ex situ* transport measurement, which is different from FeTe protection situation [2]. Figure 1(a) shows the superconducting transition of FeSe films with thickness ranging from 5 to 50 UC. As the films become thicker, the superconducting transition shifts to lower temperature. This result is more clearly in figure 1(b) where a direct correlation of $T_c$ with the film thickness is plotted. Here, $T_c$ is defined as the intersection between the linear extrapolation of the normal state and the superconducting transition, as illustrated in figure 1(a). This behavior is in sharp contrast to that of FeSe films grown on double layer graphene [28, 29], in which the interface bonding is very weak [29], and implies that interface effect might play a great role for the enhanced superconductivity of FeSe/STO. In the case of FeSe films on graphene, $T_c$ scales inversely with the thickness and no superconducting signal is detected for 1 UC FeSe film at a temperature of 2 K. The result is not surprising since the $T_c$ of a superconducting film decreases usually with thickness due to thermo and quantum fluctuations are more severe at reduced dimensionality. But the situation of FeSe/STO is different: the chemical bonding at the interface is strong and chemical bond disorders could be further diminished by post-annealing [24], which can promote the charge transfer at the interface. Previous studies indeed reveal significant charge transfer (up to 0.12 electrons per unit cell for 1 UC FeSe) [15, 25] and phonons [9] from STO substrates. Both factors can promote superconductivity according to the BCS theory. In addition, since both STS [1] and angle-resolved photoemission spectroscopy (ARPES) [5] measurements show non-superconducting behavior for the top layer of 2 UC and thicker FeSe films, the observed enhanced superconductivity of 5-50

UC films by transport should come mainly from the contribution of the first UC FeSe on STO. Therefore, one possible reason for the $T_c$ suppression with increasing thickness is that the second and above FeSe UCs may reduce the amount of charges in the first FeSe layer if we simply assume the total amount of charge transferred from STO depends only on the band offset at the interface.

To determine the crystal structure of thicker FeSe films, x-ray diffraction (XRD) measurement was conducted on 50 UC FeSe films (for thinner films, x-ray scattering signal from (103) plane is too weak to detect). The result is shown Fig. 2(a). According to the data, the film is in pure phase and oriented along c-axis since all the peaks can be indexed to (0 0 n) reflections of tetragonal FeSe. The out-of-plane lattice constants were calculated from the XRD data using (002) peaks. Figs. 2(b) and 2(c) show the (103) peaks and φ-scan, respectively. The appearance of diffraction peaks separated by 90 °C in the φ-scan indicates an in-plane alignment of the FeSe film with the STO substrate. Based on the (002) and (103) diffraction peaks, the out-of-plane lattice constant and in-plane lattice constant are calculated to be c = 0.5471 nm and a = b = 0.3881 nm, respectively. Compared to the bulk values (c=0.5518 nm and a=b=0.3765 nm) [22], it is obvious that the 50 UC FeSe films are still under tensile strain. Considering the result that the superconductivity is suppressed by strain in FeSe films grown on MgO [30], we argue that the tensile strain may also destroy the superconductivity in 2 UC and thicker FeSe films on STO but the first UC FeSe at the interface maintains the superconductivity by charge transfer and phonons from the STO substrate.

Shown in figure 3(a) are the transport properties of 5 UC FeSe films with different post-annealing time. The as-grown film exhibits an insulating behavior at low temperature. An insulator to superconductor transition appears when long time annealing was conducted. For high-temperature cuprate superconductors, insulator to superconductor transition is achieved by doping the antiferromagnetic Mott parent insulator. However, for iron-based superconductors, the parent compound is mostly bad metal. Thus, the insulator to superconductor transition demonstrated here is rather unusual and interesting. The signature of insulator to superconductor transition in 1 UC FeSe film was detected by ARPES [10]. Our result is the first direct evidence for it from transport measurement. According to the ARPES result, the transition is due to increased carrier concentration in the first UC FeSe, suggesting that enhanced electron localization or correlation at reduced dimensionality may play key role [10]. The previous STM/STS experiments show that the as-grown FeSe film is nonstoichiometric with surface Se adatoms, characterized by 0.4 eV semiconducting gap, while superconductivity is developed by post-annealing to form stoichiometric films and maximize charge transfer [25]. We found that $T_c$ is not linearly scaled with annealing time; $T_c$ of the FeSe film annealed for 55 h is just 30 K, lower than that for 36 h. The result implies that for maximum $T_c$, post-annealing must be optimized to achieve highest charge transfer. Figure 3(b) shows the corresponding Hall coefficient of 5 UC FeSe films with different annealing time. It exhibits a strong temperature-dependent behavior possibly due to the multiband structure of FeSe and the varied contribution of both electron and hole carriers from multiband, similar to their bulk courterparts [32]. More interestingly, at the low temperatures below 100 K, the dominated charge carrier is p type for the as-grown insulating FeSe film, while it is n type for all superconducting films. Additionally, the sign reversal temperature ($T_{e-h}$) for the superconducting films systematically shifts to higher temperature with increasing annealing time. Undoubtedly, long-time annealing would introduce electron charge to the system, which is crucial for the observed insulator to superconductor transition and occurrence of the superconductivity. The superconducting Cooper pairs are from the n type quasiparticles in 5 UC FeSe films on STO, which is consistent with the previous report for 1 UC and 2 UC FeSe on STO [25,26].

Figure 4(a) shows the transport properties of 5 UC FeSe film annealed for 55 h (shown in figure 3(a)) ranging from 2 to 60 K. Interestingly, with decreasing temperature, the resistance first decreases and reaches a minimum at 35 K, then, it exhibits an insulating-like behavior indicated by an upturn before the occurrence of superconducting transition and eventually approaches to zero resistance state at 9.5 K. This upturn is not observed in other samples with shorter annealing time (figure 3(a)) and possibly due to 2D localization [33]. Moreover, we should notice that the superconducting transition for this 55 h film is quite broad. Figure 4(b) exhibits the resistance as a function of temperature under magnetic field up to 15 T. In this work, the applied magnetic field is always perpendicular to the film. One can see that the resistance upturn behavior could be effectively suppressed when applying a small magnetic field below 1 T, but continuously increasing the magnetic field up to 15 T will drive the resistance upturn behavior more evidently together with shifting the superconducting transition to a lower temperature range. Since the upper critical magnetic field ($H_{C2}$) of the film is too high to be achieved, the magnetic field of 15 T cannot tune the film into a metal or an insulator state at low temperatures. But the data still show a trend for superconductor-insulator transition (SIT) or superconductor-metal transition (SMT) tuned by the applied magnetic field. Figure 4(c) shows the resistance as a function of magnetic field at the temperature of 30 K, it is apparent that a crossover from negative magnetoresistance (NMR) to linear positive magnetoresistance (PMR) occurs as continuously increasing the magnetic field. When the magnetic field is smaller than 1 T, the resistance is decreasing with increasing the magnetic field. Above this magnetic field value, the resistance increases rather linearly. Notably, the NMR in figure 4(c) corresponds to the suppression of the resistance upturn at around 30 K as shown in figure 4(b).

For comparison, figure 4(d)-4(f) show the superconducting properties of a typical 5 UC FeSe film covered by Si protection layer with annealing time around 40 hours under different magnetic fields. $T_c$ of this sample is also 30 K. Comparing with the sample annealed for 55 hours, no insulating-like behavior (figure 4(d)) and NMR (figure 4(e)) are observed in this sample since the disorder is weaker with 40 hours annealing. Moreover, the linear PMR has also been demonstrated here at the temperature of 20 K as shown in figure 4(e). The temperature dependence of the upper critical magnetic field $H_{C2}$ of this sample is shown in figure 4(f). Under the perpendicular magnetic field, $H_{C2}$ shows linear T dependence, which follows the standard linearized GL theory for "ideal" 2D superconductors, $H_{C2}^{\perp} = \frac{\Phi_0}{2\pi\xi_{GL}(0)^2}(1-\frac{T}{T_C})$, where $\Phi_0$ is the flux quantum and $\xi_{GL}(0)$ is the GL coherence length at zero temperature. Here, for the linear fitting, the middle resistance temperature of superconducting transition ($T_c^{mid}$) is used since it is more distinct to the applied magnetic field. In this way, the upper critical field $H_{C2}$ for the 5 UC FeSe can be extrapolated to be 45.0 T, which is smaller than that of 1 UC FeSe covered by FeTe protection layer [2] (64.9 T) since the $T_c$ here is much lower.

For superconducting samples, in the vicinity of the superconducting transition temperature, transport is dominated by the superconducting fluctuations. Therefore, PMR could be expected near the superconducting transition due to the suppression of superconductivity. NMR, however, is unexpected. The NMRs have ever been reported in disordered conventional superconducting thin films [34-36]. For the fractal Pb thin films on Si(111), the NMR with wide magnetoresistance terrace under perpendicular magnetic field was reported and attributed to the coexistence of two superconducting phases in this factual film [36]. In the case of granular Pb films and amorphous indium oxide thin films, a large NMR was observed on the insulating side of 2D superconductor-insulator transition and attributed to the dominance of intergranular tunneling and the Cooper pair breaking effects, respectively [34, 35]. A distinct feature in our case is that the NMR only occurs in the vicinity of superconducting transition under a relatively weak magnetic field and is correlated well with the upturn behavior shown in figure 4(a). The intensive annealing would unavoidably introduce Se vacancies at random position and even deteriorate the film quality which could be treated as scattering centers, thus making the film more inhomogeneous with rather broad superconducting transition. Therefore, one probable description for the NMR would be the 2D weak localization effects, where even a weak magnetic field can break the time-reversal symmetry and thus drive more charge carriers contributing to the conductance [37]. Besides, the observed linear magnetoresistance is also unconventional, which might represent Dirac-like structure in FeSe films [20, 38] and deserves further investigations.

**Conclusion**

In summary, we firstly report the thickness dependence of superconductivity and superconductor-insulator transition in ultrathin FeSe films grown on STO substrate by *ex situ* transport measurement. We find that $T_c$ is gradually suppressed with increasing film thickness, which is opposite to the relation in traditional superconducting films and indicates the important role from the interface effect. Considering that superconductivity is not detected by STM/STS and ARPES for 2 UC and thicker FeSe films on STO, we conclude that our observation of superconductivity in thicker FeSe films on STO by transport measurements demonstrates the unique feature of interface superconductivity, that is, only the first UC FeSe on STO is superconducting whatever the thickness is larger than 1 UC or not. As for the 5 UC FeSe films, an insulator to superconductor transition occurs as increasing the annealing time. Corresponding Hall measurements identify that the dominated carriers are holes for insulating films and electrons for superconducting films. Besides, for the FeSe film under intensive annealing, a resistance upturn around $T_c$, which presents a trend of SIT tuned by magnetic field, is observed. Simultaneously, an unexpected crossover from NMR to linear PMR is revealed as sweeping magnetic field, indicating possible 2D weak localization effect and Dirac-like structure in such FeSe films. Our findings may open the door to further understand the interface-enhanced high-temperature superconductivity and stimulate the studies in this new and rapidly developing frontier.

**Methods**

FeSe films were grown in a commercial Omicron UHV STM system combined with MBE. The base pressure of the system is better than $2\times10^{-10}$ mbar. Insulating STO substrates were used in our experiment in order to carry out transport measurement. The substrates were prepared under standard chemical and thermal treatments in order to obtain a specific $TiO_2$-terminated surface [2]. After that, the STO substrates were transferred into the UHV chamber and degased at 600 ℃ for 3 h before film deposition. FeSe films were grown by evaporating Fe (99.995%) and Se (99.9999%) with a nominal flux ratio of 1:10 from two standard Knudsen cells. During the growth, the STO substrate was held at 400 ℃ and the growth rate was maintained at 0.2 UC/min. Under the above growth condition, large-scale homogeneous FeSe films can be obtained, which is crucial for transport measurements. Prior to being transferred out of UHV for *ex situ* measurements, 20 nm thick amorphous Si capping layers were deposited on FeSe films as protection layers. The structures of thick FeSe films were further characterized by an x-ray diffractometer, a two-circle x-ray diffractometer with monochromatic Cu K$\alpha$=1.540 598 Å radiation.

**Acknowledgements**

We acknowledge Dunghai Lee for helpful discussions. This work was financially supported by the National Basic Research Program of China (Grant Nos. 2013CB934600 and 2012CB921300), the National Natural Science Foundation of China (Grant Nos. 11222434 and 11174007), and the Research Fund for the Doctoral Program of Higher Education (RFDP) of China.

Figure caption

Figure 1

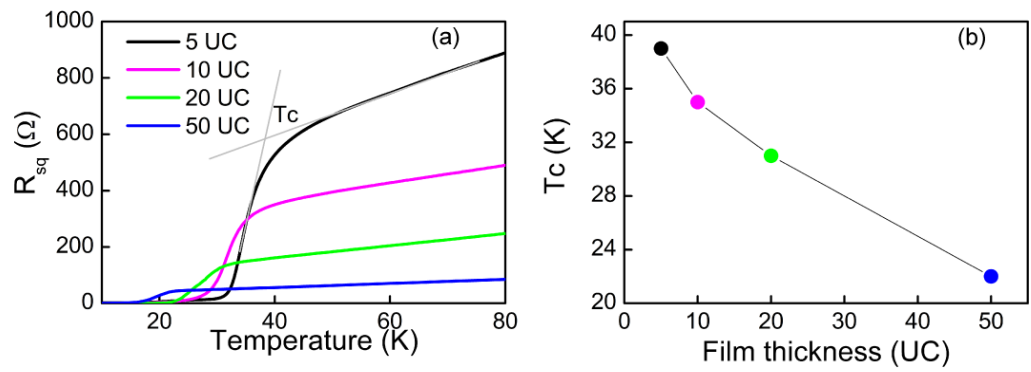

Figure 1. (a) Transport measurement of ultrathin FeSe films on STO with varied thickness. $T_c$ is defined by the grey lines. (b) $T_c$ as a function of FeSe film thickness.

Figure 2

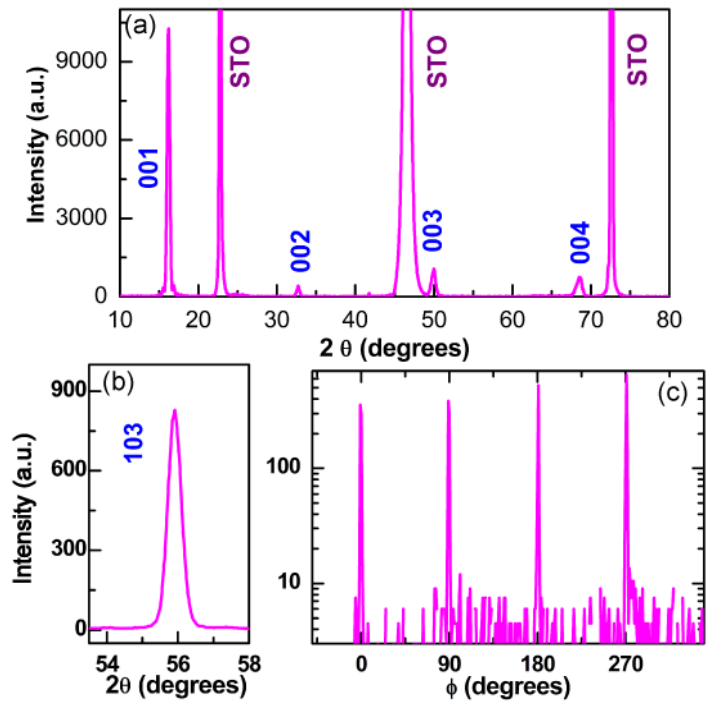

Figure 2. (a) The x-ray diffraction pattern of 50 UC FeSe film measured at room temperature, only (00n) peaks of tetragonal phase FeSe observed, suggesting strong c-axis preferred orientation. (b) The x-ray diffraction (103) peaks and its corresponding φ scan shown in (c).

Figure 3

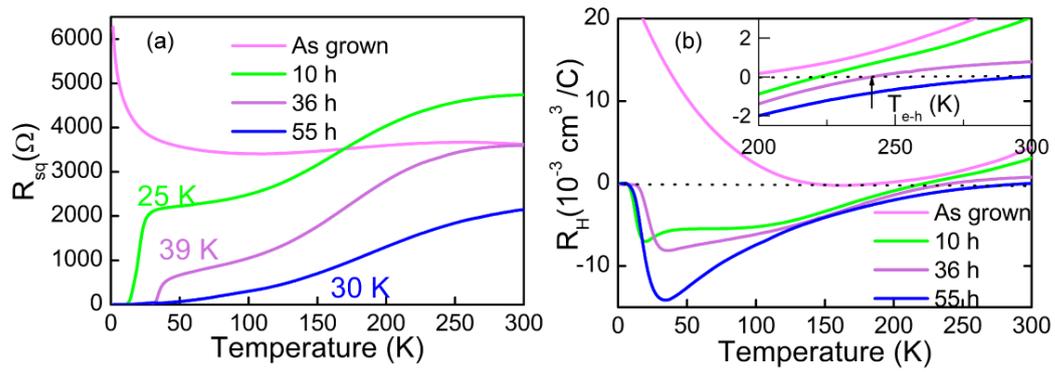

Figure 3. (a) The sheet resistance as a function of temperature for the as-grown 5 UC film and also 5 UC films annealed at 500℃ with various post-annealing time. The number marks the $T_c$ of the $R(T)$ curve with the same color. The corresponding temperature-dependent Hall coefficient is shown in (b), the inset shows the zoom-in region near $R_H = 0$, demonstrating the change of carrier type from hole to electron at $T_{e\text{-}h}$.

Figure 4

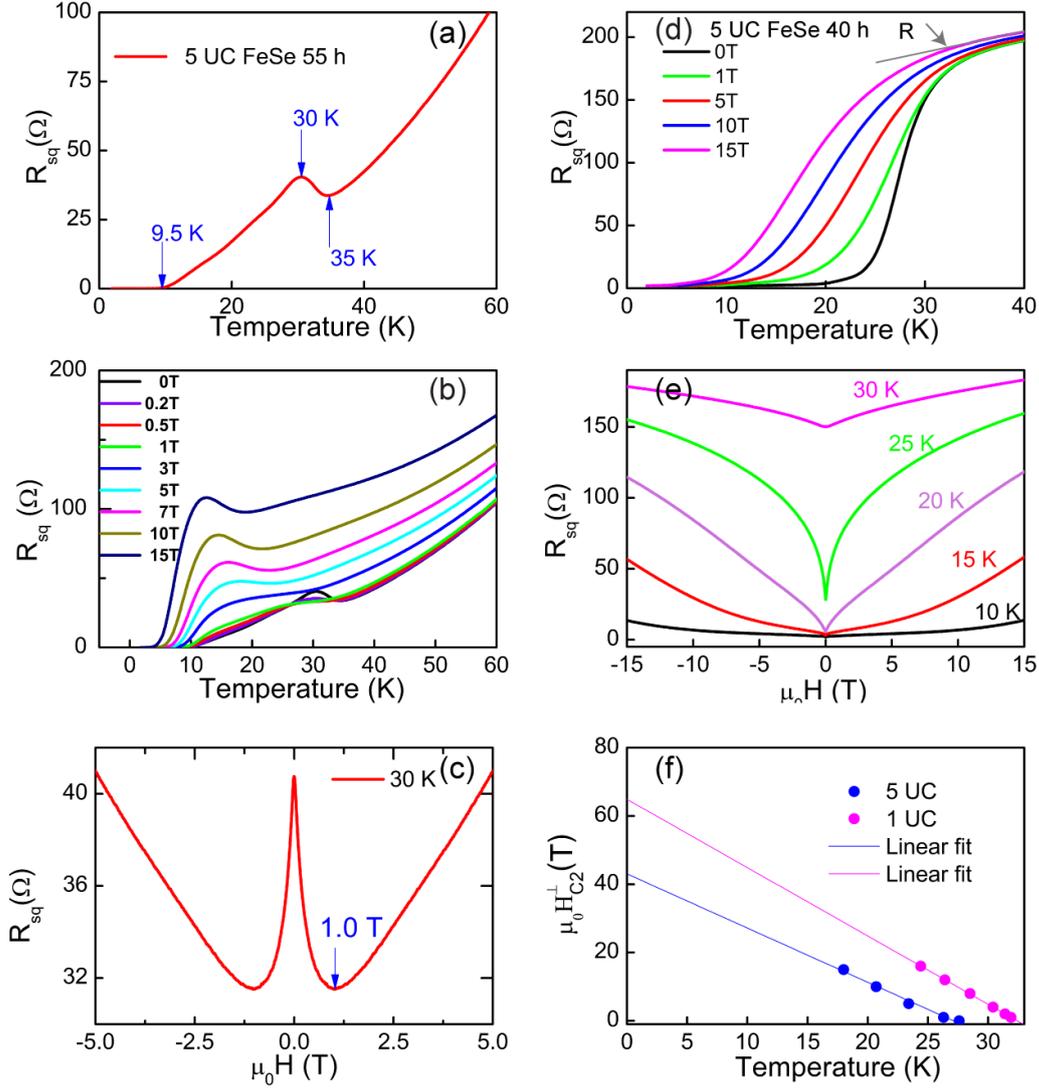

Figure 4. (a)-(c) show the eletronic properties for the 5 UC FeSe film annealed at 500℃ for 55 h with $T_c$ = 30 K. (a) The sheet resistance as a function of temperature. (b) $R(T)$ characrisitics under different magnetic fields, showing a trend of SIT tuned by magnetic field. (c) Magnetoresistance measured at 30 K as sweeping the magnetic field, demonstrating a crossover from NMR to linear PMR. (d)-(f) show the electronic properties for a typical 5 UC FeSe film annealed at 500℃ for 40 h with $T_c$ = 30 K. (d) The superconducting transition under different magnetic fields. (e) The magnetoresistance measured at varied temperatures, no NMR is demonstrated here and the MR at 20 K is rather linear. (f) Temperature-dependent of upper critical field and its linear fitting. The middle resistance temperature of superconducting transition is used here, where $R(T_c^{mid}) = R/2$ shown in (d). For comparison, the data for 1 UC FeSe sample with FeTe capping layer is also presented [2]. In our measurements, the applied magnetic field is always perpendicular to the films.